# USING NARROW BAND PHOTOMETRY TO CLASSIFY STARS AND BROWN DWARFS

A. K. Mainzer[1,2], Ian S. McLean[1], J. L. Sievers[3], E. T. Young[4]

[1]Department of Physics and Astronomy, University of California, Los Angeles, CA 90095; mclean@astro.ucla.edu

[2]Jet Propulsion Laboratory, 4800 Oak Grove Drive, Pasadena, CA 91109; amainzer@jpl.nasa.gov

[3]Canadian Institute for Theoretical Astrophysics, 60 St. George Street, Toronto, ON M5S 3H8, Canada; sievers@cita.utoronto.ca

[4]Steward Observatory, 933 N. Cherry Ave., Tucson, AZ 85721; eyoung@as.arizona.edu

## ABSTRACT

We present a new system of narrow band filters in the near infrared that can be used to classify stars and brown dwarfs. This set of four filters, spanning the H band, can be used to identify molecular features unique to brown dwarfs, such as $H_2O$ and $CH_4$. The four filters are centered at 1.495 µm ($H_2O$), 1.595 µm (continuum), 1.66 µm ($CH_4$), and 1.75 µm ($H_2O$). Using two $H_2O$ filters allows us to solve for individual objects' reddenings. This can be accomplished by constructing a color-color-color cube and rotating it until the reddening vector disappears. We created a model of predicted color-color-color values for different spectral types by integrating filter bandpass data with spectra of known stars and brown dwarfs. We validated this model by making photometric measurements of seven known L and T dwarfs, ranging from L1 – T7.5. The photometric measurements agree with the model to within ±0.1 mag, allowing us to create spectral indices for different spectral types. We can classify A through early M stars to within ±2 spectral types, late-type M and L dwarfs to within ±0.3 spectral types and T dwarfs to within ±0.1 spectral types 1-σ. Thus, we can distinguish between a T1 and a T3 dwarf. The four physical bands can be converted into two reddening-free indices, $\mu_1$ and $\mu_2$, and an extinction, $A_V$, for the individual objects. This technique, which is equivalent to extremely low resolution spectroscopy, can be used to survey large areas to provide rough spectral classifications for all the stars in the area, ranging down to the coolest brown dwarfs. It should prove particularly useful in young clusters where reddening can be high.

*Subject headings:* infrared: stars – stars: brown dwarfs – stars: formation – stars: pre-main-sequence

## 1. INTRODUCTION

Prior to the discovery of brown dwarfs (Becklin & Zuckerman 1988; Nakajima et al. 1995; Oppenheimer et. al. 1995), the main stellar sequence ended with spectral type M. Stars of spectral type M are characterized by the appearance of deep molecular absorption bands due to the formation of oxides of refractive elements, such as titanium oxide (TiO) and vanadium oxide (VO). Recent years have seen the extension of the spectral sequence to lower temperatures and substellar masses with two new types, L (with temperatures between 2500 and 1500 K) and T (with temperatures less than 1500 K). L dwarfs are characterized by the absence of TiO and VO absorption and the presence of metal hydrides and strong alkali lines in their optical spectra, and by strong water vapor absorption bands in the J, H, and K infrared wavebands (e.g. Martin et al. 1999; Kirkpatrick et al. 1999, 2000). T dwarfs are distinguished by the strengthening of water absorption bands and the onset of methane absorption in J, H, and K (e.g. Burgasser et al. 1999; Geballe et al. 2002). Until the discovery of the T dwarf Gliese 229B (Nakajima et al. 1995, Oppenheimer et al. 1995) methane absorption had only been seen in the spectra of the Jovian planets and Titan. Spectroscopic surveys, such as the NIRSPEC Brown Dwarf Spectroscopic Survey (BDSS) at the Keck Observatory (McLean et al. 2003) and others (Geballe et al. 2002), have begun to systematically characterize M, L, and T dwarfs in the infrared and produce canonical, high signal-to-noise spectra of each subtype. Figure 1 shows a number of NIRSPEC spectra from the BDSS (McLean et al. 2000). The transitions between M, L, and T dwarfs can be seen. The appearance (and strengthening) of water absorption at 1.5 μm and 1.75 μm characterizes the L dwarfs, and the addition of methane at 1.66 μm distinguishes the T dwarfs. Theory suggests that L dwarfs are a mixture of hydrogen-burning stars and sub-stellar brown dwarfs. All T dwarfs have sub-stellar masses.

In order to efficiently detect and classify faint brown dwarfs in imaging surveys, we have devised a set of narrow band near infrared filters at UCLA that select out their unique molecular features, in particular $H_2O$ and $CH_4$. Our filters completely span the H band, centered at 1.495 μm ($H_2O$), 1.595 μm (continuum), 1.66 μm ($CH_4$), and 1.75 μm ($H_2O$). All four bands are approximately 6% wide. The use of the narrow continuum filter provides strong contrast between $H_2O$ and $CH_4$ absorption bands. The first application of this technique was reported by Mainzer and McLean (2003) using only two filters and the standard H band.

One of the most difficult aspects of detecting low mass objects, particularly in young star forming regions where extinction is high due to the presence of primordial gas and dust, is distinguishing genuine cluster members from foreground or reddened background objects. Using a second $H_2O$ filter at 1.75 μm, we can distinguish between bona fide water absorption and reddening by constructing three dimensional color-color-color cubes. Reddening now appears as a vector in three dimensions. The cube can now be rotated until the reddening vector is seen face-on, making it disappear. In Section 2 we describe the development of photometric indices using the new filters. Observational verification of the technique is given in Section 3 and applications are discussed in Section 4. Our conclusions are given in Section 5.

## 2. DEVELOPMENT OF THE PHOTOMETRIC INDICES

To determine the best set of narrow band indices to use for spectral classification, a model was created based on the flux-calibrated spectra of 85 objects for which R~2000 spectra had already been obtained. These objects ranged in spectral type from B0 main sequence stars to M9.5 dwarfs, early and late L and T dwarfs, M and K giants and supergiants, and galaxies. The main sequence stars were drawn from a database created by Pickles (1998); the M dwarfs from Leggett et al. (2002), the galaxy spectra from M. Rieke (private communication, 2001), and the late M, L and T dwarf spectra from the BDSS.

Using the desired set of trial filter bandpasses as input, the model calculates the magnitudes of each object in each band by numerically integrating the total flux in that band. In addition, the model creates color-magnitude and color-color plots for each set of filter combinations. The primary characteristics of T dwarfs are strong $H_2O$ and $CH_4$ absorption bands in the near-infrared (see Figure 1). Therefore, a set of indices based on the strengths of water and methane absorption relative to a continuum band should yield the best discriminator of these objects.

Although strong $H_2O$ and $CH_4$ features are also found in the K band, it was determined that the best choice was the H band using features centered at 1.495 µm, 1.66 µm, and 1.75 µm. The locations of the two $H_2O$ filters at 1.495 µm and 1.75 µm were compromises between maximizing the stellar signature and avoiding the deepest part of the telluric absorption. For ground-based observations, the K band is less sensitive due to thermal background emission from the telescope. Although the J band could be used for the continuum filter, the lower thermal background at J band is offset by high extinction by the dust shrouding stars in young clusters. Both J and H suffer from variable OH emission in the Earth's atmosphere, but H provides the best overall sensitivity.

By examining the spectra of L and T dwarfs from the BDSS (McLean et al. 2003), it became obvious that the optimal filters for detecting water and methane should have widths of 0.1 µm, a 6% width for all four filters. The goal was to widen the bandpass as much as possible to allow maximum throughput in that absorption feature while maintaining sensitivity to the spectral structure. Figure 1 shows the BDSS spectra of several M, L and T dwarfs with the narrow-band filters overlaid. A third narrow band filter centered on 1.595 µm samples the continuum between the water and methane features and provides the best contrast between methane-bearing T dwarfs and everything else. In our previous survey of IC348 (Mainzer & McLean 2003), we did not have the narrow continuum filter. Instead, we used a broadband H filter as a proxy. This substitution resulted in reduced contrast between various spectral types. A fourth filter centered at 1.75 µm measures the edge of the 1.8 µm water band. Figure 2 shows the as-built filter transmission curves; Table 1 gives the tabulated values for each of the four filters. These curves do not include the effects of either detector response or atmospheric variations; for such a small span in wavelength across the H band, these are basically constant for InSb detectors. The filters were designed to be narrow enough to avoid the atmospheric water bands below 1.4 and above 1.8 µm. Further, since this is a differential photometric classification system, variations across the passbands affect all targets equally and are self-calibrated out.

3. OBSERVATIONS

In order to validate our spectral classification model, we observed seven L and T dwarf standards, as well as a G star (Gizis et al. 2000; Kirkpatrick et al. 2000; Reid et al. 2000; Tinney et al. 2003; Burgasser et al. 2002, Geballe et al. 2002, Persson et al. 1998). All are field objects whose spectral types are well-determined. The goal of these measurements was to determine how well the narrow band photometry agrees with the model. The observations were made using the UCLA narrow band filters installed in

FLITECAM, the First Light Camera for SOFIA (Mainzer et al. 2003), at the 3-m Shane Telescope at Mt. Hamilton on 2003 May 19-26. FLITECAM (I. McLean P.I.) is a 1 - 5 μm camera with an 8' field of view being developed at UCLA. Conditions were excellent for most of the run, with five out of seven nights photometric. Seeing was typically 0.9 – 1". Typical integration times for all sources were 27 minutes in each band, resulting in a 3-$\sigma$ sensitivity of 19.2 mag.

| Target | Spectral Type | H Band Magnitude | [1.495] | [1.595] | [1.66] | [1.75] |
| --- | --- | --- | --- | --- | --- | --- |
| 2MASS1217111-031113 | T7.5 | 15.79 | 16.04 | 14.98 | 16.46 | 16.42 |
| 2MASS13004255+1912354 | L1 | 12.07 | 12.50 | 12.14 | 11.91 | 11.98 |
| 2MASS1507476-1627386 | L5 | 11.90 | 12.31 | 11.76 | 11.61 | 11.62 |
| 2MASS15232263+3014562 | L8 | 15.00 | 15.52 | 14.86 | 14.70 | 14.60 |
| 2MASS1553022+153236 | T7 | 15.92 | 16.18 | 15.17 | 16.34 | 16.96 |
| SDSS1750+1759 | T3.5 | 15.97 | 16.35 | 15.61 | 15.83 | 16.05 |
| 2MASS2254189+312349 | T5 | 14.83 | 15.21 | 14.38 | 14.65 | 14.88 |
| SJ9150 | G | 11.31 | 11.25 | 11.26 | 11.22 | 11.12 |

Table 2: The G, L, and T standards observed with the custom narrow band filters. Zero point for [1.595] were determined by comparing objects in the same field of view with known H band magnitudes (and spectral types determined by using 2MASS colors) to the [1.595] measurements. Zero points for the remaining colors were determined by setting an A star equal to zero color, then determining the spectral type of background objects in each field using their 2MASS colors and setting their narrow band colors appropriately. [1.595] magnitudes for all objects are accurate to ±0.03 mag; [1.495], [1.66], and [1.75] magnitudes are accurate to ±0.07 mag.

The targets, their H band magnitudes, and their FLITECAM-measured narrow band magnitudes are listed in Table 2. We can now directly compare the photometric data for these sources with the passband synthesis model. Since we have four bands, we can construct a color-color-color cube. Figure 3 depicts the results of the model, with stars ranging from O – T8, for both unreddened and reddened with an $A_V$=40 reddening law applied (Rieke & Lebofsky 1985). Figure 4 shows the actual photometric data. Since these are all nearby field objects, they are assumed to have zero reddening. The dashed line in both cubes represents the reddening vector for $A_V$ = 40. The cubes can then be rotated so that we are looking down the reddening vector, making it disappear (Figures 5 and 6). In this way, we can solve for reddening for each individual object. This technique can be useful in young star forming regions, where extinction can cause confusion between a reddened background main sequence star and a water- or methane-bearing young cluster member. The advantage of this technique is that one is not required to make any assumptions or models of average reddening, and it definitively identifies and classifies individual candidates. Using a photometric classification technique can result in great savings in observing time when compared to spectroscopy; indeed, for the faintest targets, accurate spectroscopy may not be possible.

The four physical magnitudes can be transformed into a set of two reddening-free indices, $\mu_1$ and $\mu_2$, from which we can also derive the visual extinction, $A_V$. This is equivalent to finding the two-dimensional projection of the three-dimensional color-color-color cube once it has been rotated to remove reddening. It is important to note that these reddening-free indices are derived using the Rieke-Lebofsky reddening law interpolated over the narrow band filter bandpasses. These indices were calculated by using a least-squares fit to the vector defined by examining the difference between reddened and unreddened bandpass model colors for spectral types B - L. The equations to transform between the four colors and the new indices are

$$\mu_1 = 0.92 (H_B - H_A) - 0.18 (H_B - H_C) + 0.35 (H_B - H_D)$$

$$\mu_2 = 0.26 (H_B - H_A) + 0.94 (H_B - H_C) - 0.20 (H_B - H_D)$$

$$A_V = 2.95 (H_B - H_A) + 5.13 (H_B - H_C) + 27.51 (H_B - H_D) - 3.37$$

where $H_A = [1.495]$ mag, $H_B = [1.595]$ mag, $H_C = [1.66]$ mag, $H_D = [1.75]$ mag.

The reddening-free indices $\mu_1$ and $\mu_2$ were defined such that the main sequence lies along the $\mu_1$ axis. The agreement between the data and the model can be most easily seen by comparing the reddening-free indices of the model and the actual data. Figures 7 and 8 show the comparison between data and model in $\mu_1$ and $\mu_2$ reddening-free color indices. It can be seen from these figures that these indices are most sensitive to L and T dwarfs, with the latest-type T dwarfs exhibiting an ~2 mag depletion in the [1.595] – [1.75] colors. The uncertainty in the derived reddening, $A_V$, will be ±2.5 mag 1-$\sigma$ for a photometric accuracy of ±0.05 mag in all four bands. Note that the extinction determination only applies precisely to main sequence stars through L dwarfs. The index will still yield an estimate of the extinction for T dwarfs, but with diminished accuracy due to the steep structure of T dwarf absorption features, which can shift the apparent band centers. The amount of inaccuracy expected for the T dwarf sequence reddening is approximately 0.2-0.3 mag, resulting in an extinction uncertainty of 6 – 8 mag.

Table 3 gives the typical color-color-color indices for various spectral types ranging from main sequence objects through T8 as calculated from the bandpass model. Table 4 gives the color-color-color indices derived from the as-measured narrow band photometry. These are the tabulated values of Figures 3 and 4. Table 5 gives the reddening-free indices for the various spectral types as predicted by the bandpass model. Table 6 gives the reddening-free indices for the various spectral types derived from the as-measured narrow band photometry. These are the tabulated values of Figures 7 and 8.

The advantage of using FLITECAM to observe brown dwarf standards is that its wide field of view captures many other objects along with the target. This allows us to perform relative photometry between the target and other background stars. For background stars which have known 2MASS colors, it is possible to estimate their spectral types (Bessell & Brett, 1988). Thus, we can use these background stars to calibrate the color-color-color cubes for each field. This was done by making standard (J-H) vs. (H-K) color-color plots from 2MASS data for the background stars, then choosing those objects which appeared to be the least reddened on this color-color plot to serve as calibrators. In each field, there were typically only between one and three such stars available. Thus, the zero point errors were those associated with few stars. These unreddened background stars were assigned a spectral type based on their JHK colors using Tables 2 and 3 of Bessell & Brett (1998). The [1.595] zero point for a given field was set by equating the [1.595] magnitudes of the calibrator stars with their 2MASS H band magnitudes. Since the calibrator stars were chosen to be F – K stars, the H band magnitude is a reasonable representation of the continuum feature at 1.595 µm. Zero points of the remaining three colors were set by adjusting the calibrator stars' narrow band magnitudes to match their 2MASS-derived spectral types such that an A star is defined to have zero color ($H_A = H_B = H_C = H_D = 0$). By assigning these background objects with known spectral types to their appropriate color values, we can calibrate the photometry of the brown dwarfs. This makes it possible to combine fields measured on different nights in variable weather conditions into a single color-color-color cube. Comparison between 2MASS broadband H and [1.595] magnitudes showed agreement to within ±0.04 mag for the calibrator stars. This error is derived as follows: The 2MASS magnitudes are accurate to ~±0.03 mag; this is added in quadrature with the FLITECAM photometry of individual objects within a single field, which is accurate to ±0.03 mag. The errors for the [1.495], [1.66], and [1.75] bands were measured to be ±0.07 mag. Although photometry was measured to within ±0.03 mag in all of these bands, the uncertainty in the assignment of zero points using the 2MASS colors of background objects is the dominant error source. This was derived by calculating the distribution in the $\mu_2$ axis when plotting all the background objects for each field in $\mu_1$ vs. $\mu_2$ space. For A through early M stars, the variation in the $\mu_2$ axis should have been ±0.01 mag according to the bandpass model (see Table 5). Therefore, any variations greater than this are attributable to a combination of photometric error and zero point error. It is important to note that the zero point errors of the other three bands will be reduced greatly by performing relative photometry between objects in the same field of view, e.g. a single pointing in a young cluster.

The advantage of this system is that it becomes more sensitive and better able to determine spectral type as object temperature decreases; thus, it is most sensitive to the coolest objects. Based on these results, we now have a system that can classify A through early M stars to within ±2 spectral types, late-type M and L dwarfs to within ±0.3 spectral types 1-$\sigma$, and T dwarfs to within ±0.1 types 1-$\sigma$ for objects measured with SNR = 20 (±0.05 mag photometry). Thus, a G star's spectral type would appear to be between A to early M; an M9 dwarf would appear to be between M6 – L3; and a T6 would appear to be between T5 – T7. Since $\mu_1$ and $\mu_2$ both use all four fluxes, their errors are correlated and form an error ellipse. This ellipse is plotted in the corner of Figure 7 for the case of all bands having ±0.05 mag errors. The direction of the axes of the ellipse is given by the eigenvectors of the covariance matrix of $\mu_1$ and $\mu_2$ (assuming independent, equal photometric errors in all four bands). The magnitudes of the semimajor and semiminor error ellipse axes are given by the square root of the eigenvalues of the covariance matrix multiplied by the per-band photometric errors. For ±0.05 mag errors in all four bands, the projection of the error ellipse along the B – L dwarf sequence is 0.07 mag. Since the difference between early and late L dwarfs spans 0.3 mag (see Table 5), we can determine spectral type to within 1/4 of a spectral type, or ~±0.3 spectral types 1-$\sigma$. For T dwarfs, the error ellipse's semimajor axis lies roughly along the same direction as the T dwarf sequence in Figure 7. Thus, ±0.05 mag errors in all four bands yield an error of ±0.08 mag in the semimajor axis direction. Since the T dwarf sequence spans a length of ~1.2 mag in Figure 7, we can discern a difference of approximately ±0.1 spectral types 1-$\sigma$.

| Spectral Type | [1.595] – [1.495] | [1.595] – [1.66] | [1.595] – [1.75] |
|---|---|---|---|
| B0 – A5 | 0.00 to 0.15 | -0.03 to 0.05 | -0.07 to 0.12 |
| F0 – G8 | 0.00 to 0.06 | 0.02 to 0.05 | 0.03 to 0.013 |
| K – M3 | -0.11 to -0.05 | 0.05 to 0.09 | 0.09 to 0.17 |
| M4 – M5 | -0.15 to -0.09 | 0.08 to 0.11 | 0.09 to 0.17 |
| M6 – M7 | -0.27 to -0.13 | 0.07 to 0.13 | 0.09 to 0.17 |
| M8 – M8.5 | -0.27 to -0.23 | 0.09 to 0.11 | 0.11 to 0.13 |
| M9 – M9.5 | -0.35 to -0.25 | 0.10 to 0.15 | 0.14 to 0.19 |
| L0 – L1 | -0.35 to -0.30 | 0.12 to 0.13 | 0.14 to 0.15 |
| L2 | -0.36 to -0.59 | 0.13 to 0.25 | 0.16 to 0.28 |
| L3 – L4 | -0.54 to -0.48 | 0.15 to 0.16 | 0.09 to 0.21 |
| L5 – L7 | -0.61 to -0.53 | 0.13 to 0.18 | 0.00 to 0.20 |
| L7.5 – L8 | -0.62 to -0.57 | 0.14 to 0.15 | 0.05 to 0.14 |
| T1 – T2 | -0.87 to -0.79 | 0.04 to 0.11 | -0.27 to -0.17 |
| T5 | -0.88 to -0.87 | -0.33 to -0.23 | -0.67 to -0.59 |
| T6 | -0.97 to -0.81 | -0.61 to -0.57 | -1.00 to -0.92 |
| T6.5 | -0.75 | -0.91 | -1.21 |
| T8 | -1.09 | -1.14 | -1.59 |

Table 3: Object color indices as predicted by the integrated bandpass model.

| Field Objects' Spectral Types | [1.595] – [1.495] | [1.595] – [1.66] | [1.595] – [1.75] |
|---|---|---|---|
| G | 0.01 | 0.04 | 0.13 |
| L1 | -0.36 | 0.23 | 0.16 |
| L5 | -0.55 | 0.15 | 0.14 |
| L8 | -0.66 | 0.16 | 0.26 |
| T3.5 | -0.73 | -0.22 | -0.44 |
| T5 | -0.83 | -0.27 | -0.50 |
| T7 | -1.01 | -1.17 | -1.79 |
| T7.5 | -1.06 | -1.48 | -1.44 |

Table 4: The as-measured color indices of the objects.

| Spectral Type | $\mu_1$ | $\mu_2$ |
|---|---|---|
| B0 – A5 | 0.00 to 0.11 | 0.00 to 0.03 |
| F0 – G8 | 0.03 to 0.06 | 0.00 to 0.03 |
| K – M3 | -0.07 to 0.00 | 0.00 to 0.03 |
| M4 – M5 | -0.11 to -0.07 | 0.00 to 0.03 |
| M6 – M7 | -0.21 to -0.08 | 0.00 to 0.04 |
| M8 – M8.5 | -0.23 to -0.18 | 0.00 to 0.02 |
| M9 - M9.5 | -0.28 to -0.19 | 0.00 to 0.02 |
| L0 – L1 | -0.30 to -0.25 | 0.00 to 0.02 |
| L2 | -0.49 to -0.29 | 0.00 to 0.02 |
| L3 – L4 | -0.47 to -0.40 | -0.02 to 0.00 |
| L5 – L7 | -0.57 to -0.47 | -0.03 to -0.01 |
| L7.5 – L8 | -0.58 to -0.50 | -0.04 to -0.03 |
| T1 – T2 | -0.90 to -0.81 | -0.13 to -0.7 |
| T5 | -0.98 to -0.97 | -0.40 to -0.33 |
| T6 | -1.13 to -0.96 | -0.63 to -0.57 |
| T6.5 | -0.95 | -0.81 |
| T8 | -1.35 | -1.04 |

Table 5: The reddening-free indices predicted for various spectral types computed from the integrated bandpass model.

| Field Objects' Spectral Types | $\mu_1$ | $\mu_2$ |
|---|---|---|
| G | 0.05 | 0.01 |
| L1 | -0.32 | 0.09 |
| L5 | -0.48 | -0.03 |
| L8 | -0.55 | -0.07 |
| T3.5 | -0.79 | -0.31 |
| T5 | -0.89 | -0.37 |
| T7 | -1.35 | -1.00 |
| T7.5 | -1.21 | -1.38 |

Table 6: The as-measured reddening-free indices for the objects.

4. FUTURE APPLICATIONS

The advantage of this technique is that it can be used to obtain spectral classifications and reddenings for individual objects detected in large-area surveys. This technique is particularly well-suited to surveys of young clusters, where the objects' youth and correspondingly increased brightness makes it possible to detect much lower-mass objects at farther distances and where extinction is a significant factor. It is possible that changes in the objects' spectra due to decreased gravity may alter their spectral features (c.f. Lucas & Roche 2000), producing changes in the observed narrow band colors. However, in our previous survey of IC348 (Mainzer & McLean 2003), although we found an average ~0.2 mag difference between field objects and young M dwarfs, their relative narrow band colors still served as an effective means of obtaining their spectral types.  Future identifications of young low-mass objects will determine how gravity and age influence narrow band colors.

Narrow band photometry can also be used as an effective means to find field brown dwarfs, as well as serve as a useful follow-up method for objects found via 2MASS or SIRTF.

5. CONCLUSIONS

We have designed a new set of four near infrared narrow band filters spanning the H band that can be used to simultaneously identify and classify stars and brown dwarfs. By breaking the H band into four bands centered on characteristic molecular features of cool stars and brown dwarfs, we are essentially performing extremely low resolution spectroscopy.  The advantage of narrow band photometry is that an estimate of reddening-free spectral type is provided for all objects in the field of view, avoiding the costly process of obtaining spectra for each individual object.  By using four filters, we now have a means to break the degeneracy between increased water absorption and reddening.  We have constructed a model using actual spectra and the as-measured narrow band filter transmission curves.  By observing a number of objects whose spectral types were previously known with the narrow band filters, we have confirmed the accuracy of the system.  We have devised a methodology to obtain the spectral types of various background stars in the same field of view as the targets from 2MASS, thus allowing us to combine observations made on different nights under variable weather conditions.  For objects measured with SNR = 20 ($\pm 0.05$ mag photometry), we can classify A through early M stars to within $\pm 2$ spectral types, late-type M and L dwarfs to within $\pm 0.3$ spectral types, and T dwarfs to within $\pm 0.1$ types 1-$\sigma$. The two new reddening-free indices plus extinction we have derived from the narrow band colors, $\mu_1$, $\mu_2$, and $A_V$, make this system particularly suitable for use in young clusters where extinction is high.  The use of narrow band photometry can provide an efficient means of obtaining spectral types of a great number of objects simultaneously, both in young clusters and in the field.

The authors wish to thank the entire staff of the Infrared Lab at UCLA for their outstanding support in the development and deployment of FLITECAM, as well as the staff of Lick Observatory. We also thank James Graham for graciously combining observing time with us at Lick Observatory and Mark McGovern for help with NIRSPEC BDSS data. AKM was supported by a NASA Graduate Student Research Fellowship.

This research has made use of the NASA/IPAC Infrared Science Archive, which is operated by the Jet Propulsion Laboratory, California Institute of Technology, under contract with the National Aeronautics and Space Administration.  Portions of this research were performed at the Jet Propulsion Laboratory.

Figure 1: A comparison of brown dwarf spectral types, ranging from M6.5 through T8. The FLITECAM custom narrow band filters are superimposed. (Spectra courtesy of the Brown Dwarf Spectroscopic Survey.)

Figure 2: The transmission curves of the four custom UCLA narrow band filters.

Figure 3: The model color-color-color cube, derived by integrating actual spectral over the as-measured filter bandpasses, shown with sources both reddened (diamonds, darker colors) and unreddened (circles, brighter colors). The color code is as follows: blue = main sequence stars A – K; red = M dwarfs and giants; green = L dwarfs; teal = T dwarfs. The red dashed line represents a reddening of $A_V = 40$.

Figure 4: The color-color-color cube for the as-measured photometric data. The color code is as follows: blue = G star; red = probable A – M stars; green = L dwarfs; teal = T dwarfs. The red dashed line represents a reddening of $A_V = 40$.

Figure 5: The model color-color-color cube, rotated to make the reddening vector disappear. The projection of the stellar sequence along this direction results in two reddening-free indices, $\mu_1$ and $\mu_2$. The color code is the same as Figure 3.

Figure 6: The color-color-color cube for the as-measured photometric data, rotated to make the reddening vector disappear. The projection of the stellar sequence along this direction results in two reddening-free indices, $\mu_1$ and $\mu_2$. The color code is the same as Figure 4.

Figure 7: The model, with colors converted to the reddening-free indices $\mu_1$ and $\mu_2$. The color code is the same as Figure 3. The error ellipse for ±0.05 mag photometry in all four bands is shown in blue at the lower right of the plot; note that it resembles a line, since most of the error occurs in the direction of the semimajor axis.

Figure 8: The as-measured data, with colors converted to the reddening-free indices $\mu_1$ and $\mu_2$. The color code is the same as Figure 4. The error ellipse for ±0.05 mag photometry in all four bands is shown in blue at the lower right of the plot; note that it resembles a line, since most of the error occurs in the direction of the semimajor axis.

| $\lambda$ (μm) | [1.66] Transmission | $\lambda$ (μm) | [1.495] Transmission | $\lambda$ (μm) | [1.595] Transmission | $\lambda$ (μm) | [1.75] Transmission |
|---|---|---|---|---|---|---|---|
| 1.550 | 0.00 | 1.390 | 0 | 1.505 | 0 | 1.650 | 0 |
| 1.570 | 0.00 | 1.395 | 0 | 1.510 | 0.01 | 1.655 | 0.01 |
| 1.575 | 0.01 | 1.400 | 0.01 | 1.515 | 0.02 | 1.660 | 0.01 |
| 1.580 | 0.01 | 1.405 | 0.02 | 1.520 | 0.03 | 1.665 | 0.02 |
| 1.585 | 0.02 | 1.410 | 0.02 | 1.525 | 0.05 | 1.670 | 0.03 |
| 1.590 | 0.05 | 1.415 | 0.03 | 1.530 | 0.08 | 1.675 | 0.05 |
| 1.595 | 0.10 | 1.420 | 0.05 | 1.535 | 0.15 | 1.680 | 0.15 |
| 1.600 | 0.22 | 1.425 | 0.08 | 1.540 | 0.25 | 1.685 | 0.30 |
| 1.605 | 0.42 | 1.430 | 0.10 | 1.545 | 0.35 | 1.690 | 0.45 |
| 1.610 | 0.69 | 1.435 | 0.16 | 1.550 | 0.45 | 1.695 | 0.65 |
| 1.615 | 0.77 | 1.440 | 0.24 | 1.555 | 0.55 | 1.700 | 0.69 |
| 1.620 | 0.78 | 1.445 | 0.38 | 1.560 | 0.63 | 1.705 | 0.69 |
| 1.625 | 0.75 | 1.450 | 0.50 | 1.565 | 0.70 | 1.710 | 0.70 |
| 1.630 | 0.73 | 1.455 | 0.54 | 1.570 | 0.75 | 1.715 | 0.70 |
| 1.635 | 0.72 | 1.460 | 0.58 | 1.575 | 0.78 | 1.720 | 0.66 |
| 1.640 | 0.72 | 1.465 | 0.60 | 1.580 | 0.78 | 1.725 | 0.63 |
| 1.645 | 0.74 | 1.470 | 0.63 | 1.585 | 0.77 | 1.730 | 0.64 |
| 1.650 | 0.77 | 1.475 | 0.65 | 1.590 | 0.78 | 1.735 | 0.67 |
| 1.655 | 0.79 | 1.480 | 0.66 | 1.595 | 0.77 | 1.740 | 0.71 |
| 1.660 | 0.81 | 1.485 | 0.66 | 1.600 | 0.77 | 1.745 | 0.75 |
| 1.665 | 0.82 | 1.490 | 0.66 | 1.605 | 0.76 | 1.750 | 0.78 |
| 1.670 | 0.82 | 1.495 | 0.67 | 1.610 | 0.75 | 1.755 | 0.80 |
| 1.675 | 0.82 | 1.500 | 0.66 | 1.615 | 0.77 | 1.760 | 0.79 |
| 1.680 | 0.80 | 1.505 | 0.65 | 1.620 | 0.78 | 1.765 | 0.78 |
| 1.685 | 0.80 | 1.510 | 0.64 | 1.625 | 0.80 | 1.770 | 0.78 |
| 1.690 | 0.80 | 1.515 | 0.62 | 1.630 | 0.79 | 1.775 | 0.77 |
| 1.695 | 0.70 | 1.520 | 0.59 | 1.635 | 0.78 | 1.780 | 0.69 |
| 1.700 | 0.50 | 1.525 | 0.55 | 1.640 | 0.65 | 1.785 | 0.55 |
| 1.705 | 0.26 | 1.530 | 0.48 | 1.645 | 0.35 | 1.790 | 0.40 |
| 1.710 | 0.15 | 1.535 | 0.43 | 1.650 | 0.15 | 1.795 | 0.29 |
| 1.715 | 0.08 | 1.540 | 0.35 | 1.660 | 0.12 | 1.800 | 0.10 |
| 1.720 | 0.04 | 1.545 | 0.28 | 1.665 | 0.08 | 1.805 | 0.07 |
| 1.725 | 0.02 | 1.550 | 0.17 | 1.670 | 0.03 | 1.810 | 0.05 |
| 1.730 | 0.02 | 1.555 | 0.13 | 1.675 | 0.02 | 1.815 | 0.03 |
| 1.735 | 0.01 | 1.560 | 0.08 | 1.680 | 0.01 | 1.820 | 0.02 |
| 1.740 | 0.00 | 1.565 | 0.07 | 1.685 | 0 | 1.825 | 0.01 |
| 1.745 | 0.00 | 1.570 | 0.05 | 1.690 | 0 | 1.830 | 0 |

Table 1: The tabulated throughput values for the four UCLA narrow band filters.

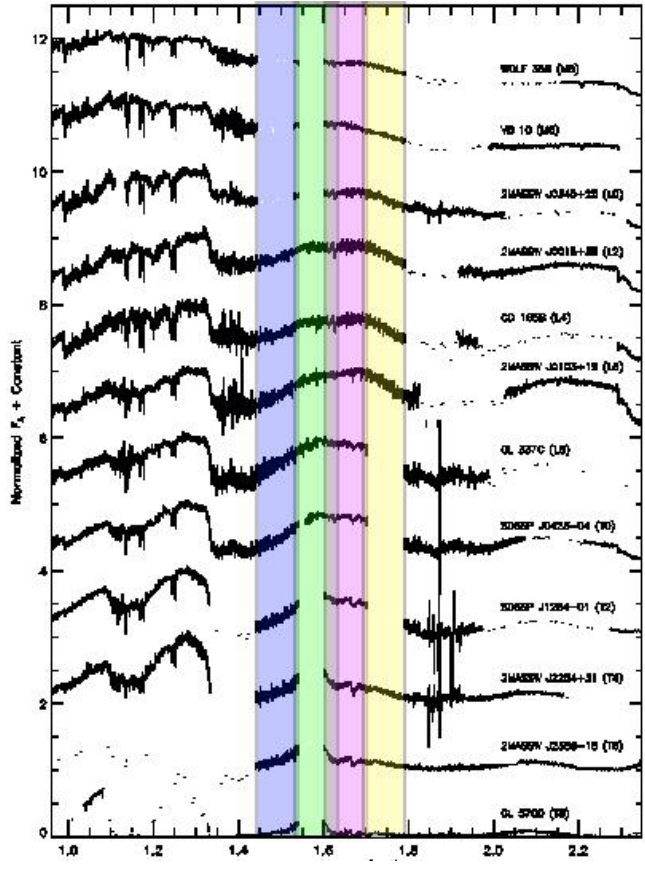

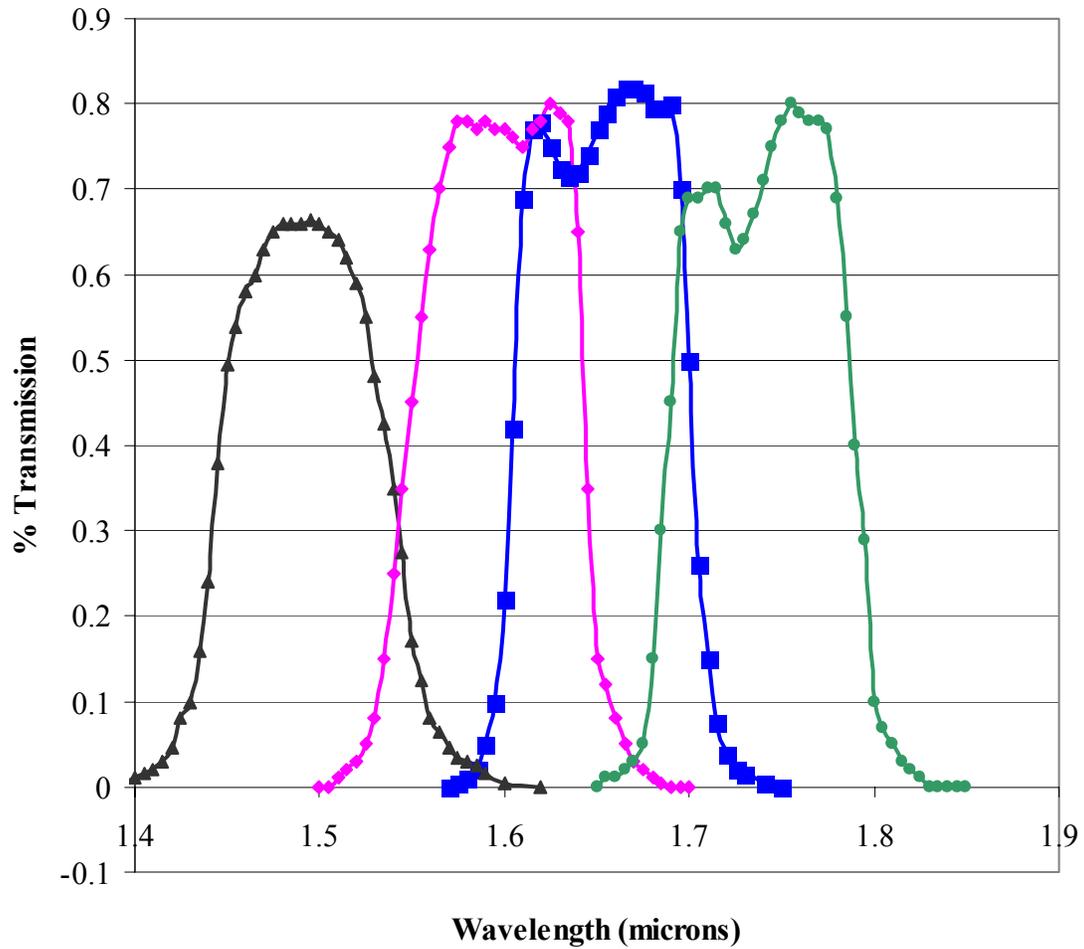

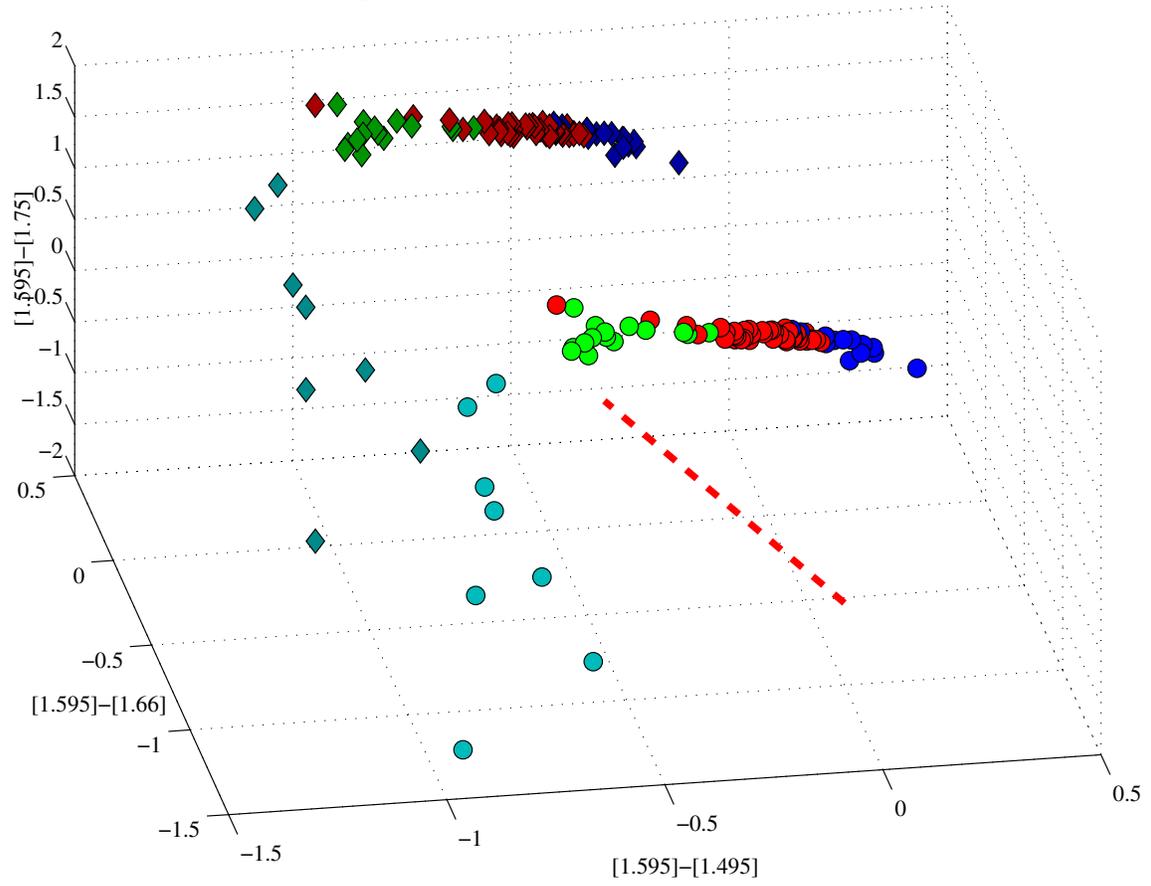

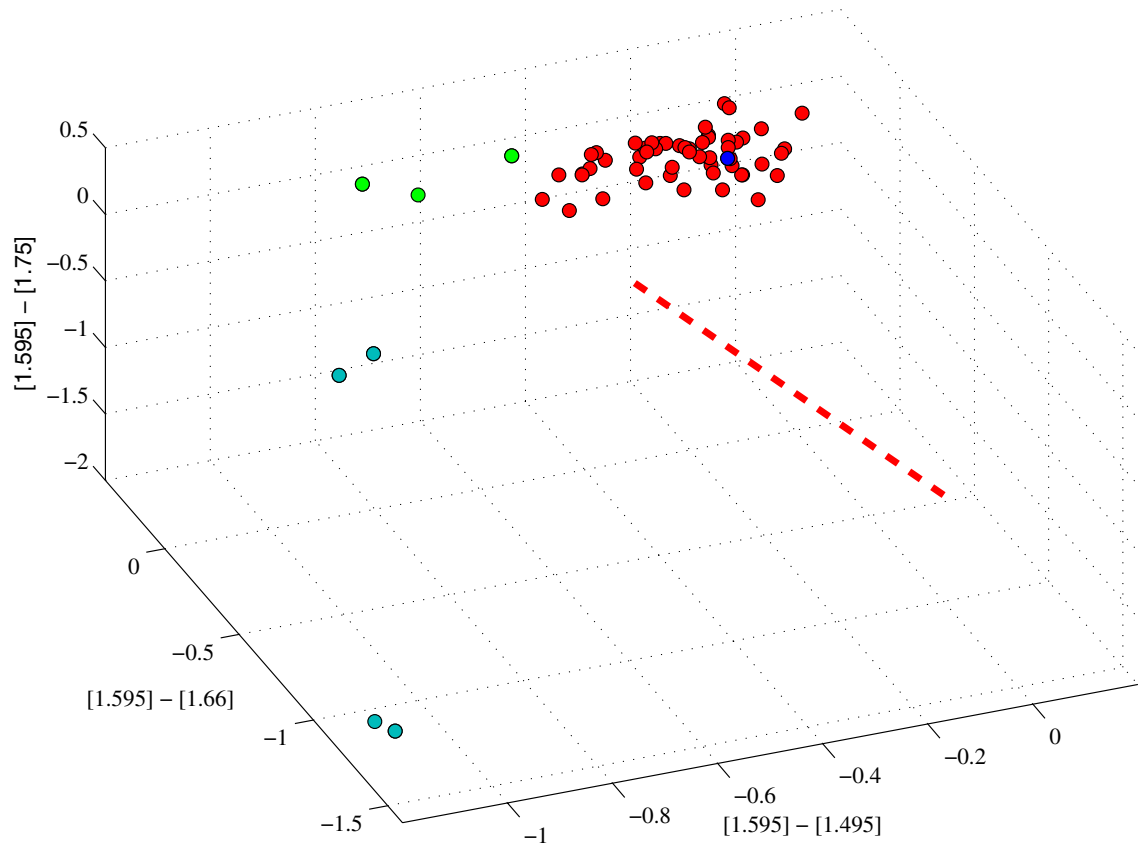

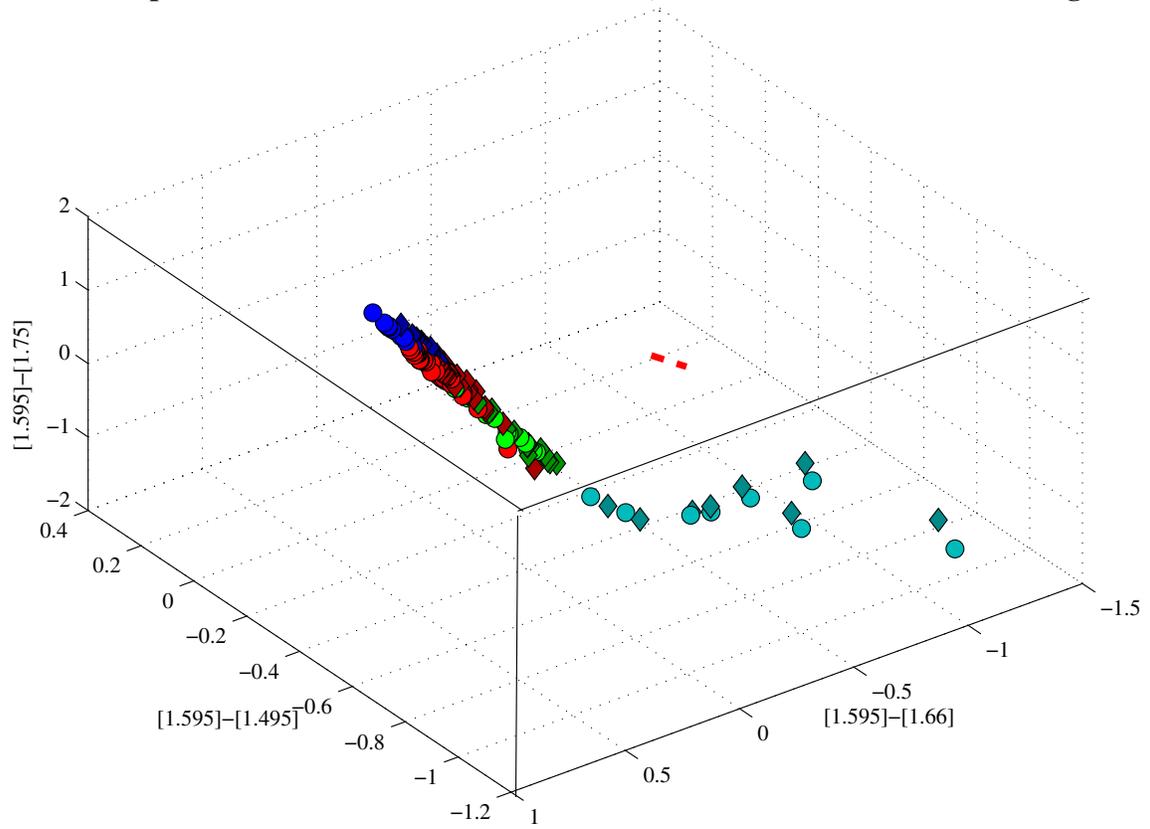

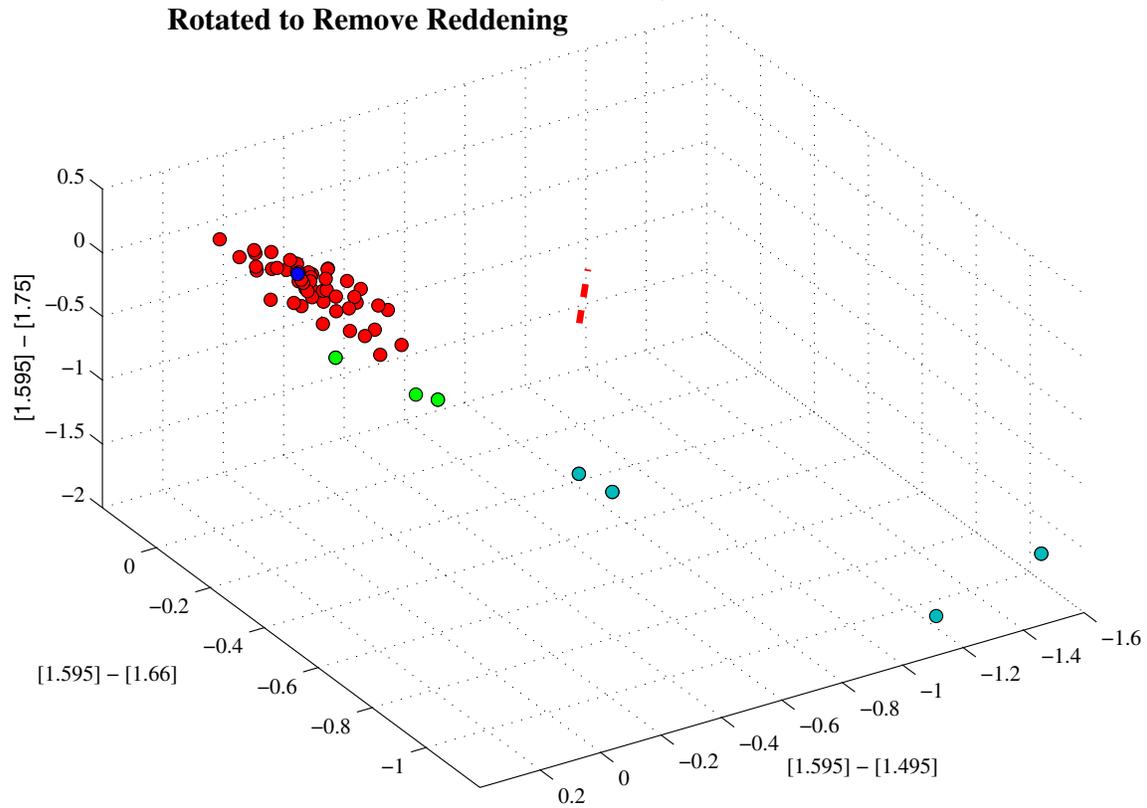

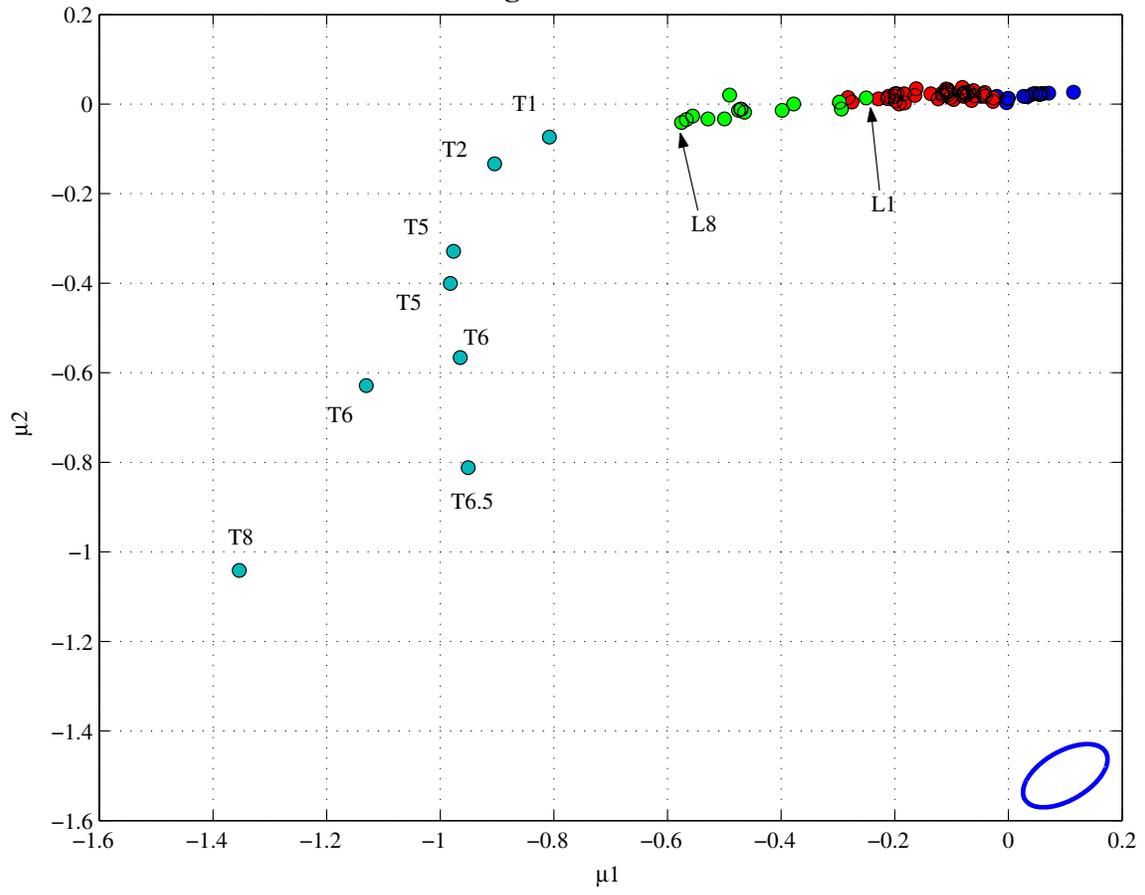

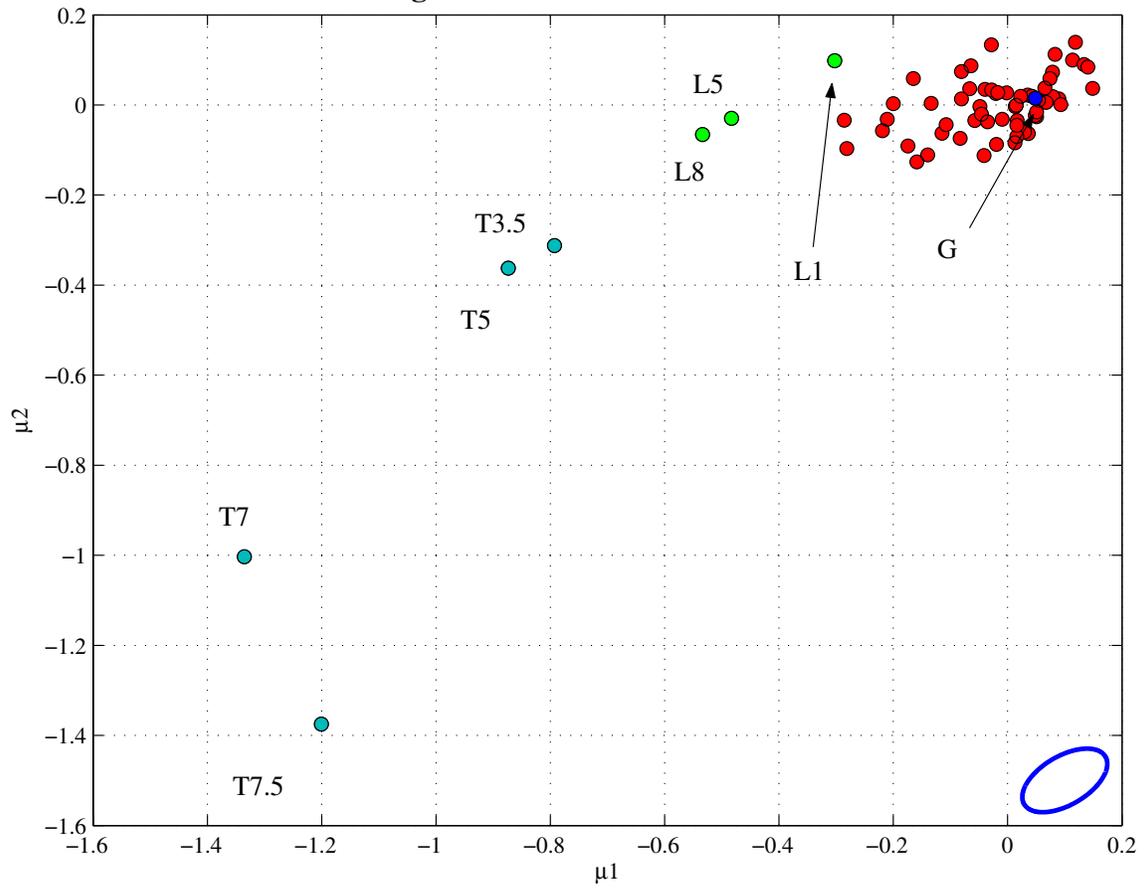